\documentclass[12pt,a4paper]{article}
\usepackage[utf8]{inputenc}
\usepackage{setspace}
\usepackage{graphicx}

\usepackage[round, sort, numbers, authoryear]{natbib}
\usepackage[english]{babel}
\renewcommand{\bibname}{References}
\usepackage{amsmath}
\usepackage{amsfonts}
\usepackage{amssymb}
\usepackage{fixltx2e,fix-cm}
\usepackage{subfigure}
\usepackage{makeidx}
\usepackage{multicol}
\usepackage{mathrsfs}
\usepackage{tensor}
\usepackage{hyperref}
\usepackage{authblk}
\usepackage{bm}
\graphicspath{{./graphics/}}
\DeclareGraphicsExtensions{.pdf,.png,.jpg}
\usepackage{amssymb}
\usepackage{amsmath}
\usepackage{graphicx}

\usepackage{bm}
\graphicspath{{./graphics/}}
\DeclareGraphicsExtensions{.pdf,.png,.jpg}

\newcommand{\ncom}{\newcommand}
\ncom{\beqn}{\begin{eqnarray*}}
\ncom{\eeqn}{\end{eqnarray*}}

\ncom{\beq}{\begin{eqnarray}}
\ncom{\eeq}{\end{eqnarray}}

\newtheorem{thm}{Theorem}[section]
\newtheorem{lemma}[thm]{Lemma}
\newtheorem{cor}[thm]{Corollary}
\newtheorem{pro}[thm]{Proposition}
\newtheorem{example}[thm]{Example}
\newtheorem{definition}[thm]{Definition}
\newtheorem{remark}[thm]{Remark}
\ncom{\et}{\end{thm}}
\ncom{\bl}{\begin{lemma}}
\ncom{\el}{\end{lemma}}
\ncom{\bco}{\begin{cor}}
\ncom{\eco}{\end{cor}}
\ncom{\bp}{\begin{pro}}
\ncom{\ep}{\end{pro}}
\ncom{\bex}{\begin{example}}
\ncom{\eex}{\end{example}}
\ncom{\bd}{\begin{definition}}
\ncom{\ed}{\end{definition}}
\ncom{\brm}{\begin{remark}}
\ncom{\erm}{\end{remark}}

\def\sI{\mathscr{I}}
\def\sR{\mathscr{R}}

\def\sM{\mathscr{M}}

\ncom{\ts}{\tensor}

\makeindex
\begin{document}
\vspace*{1cm}
\newcommand*{\email}[1]{%
    \normalsize\href{mailto:#1}{#1}\par
    }
\title{Gravitational waves-a new window to Cosmos}
\bigskip
\author{A.R.Prasanna\footnote{\textbf{Vaidya - Raichaudhuri award lecture}, delivered on 22.September 2016, Department of Physics, Lucknow University}}
\affil{Physical Research Laboratory, Navarangpura, Ahmedabad 380009,India \\ \email{prasanna@prl.res.in}.}

\maketitle

\begin{abstract}
With the detection of Gravitational waves just about an year ago  Einstein`s general theory of relativity- 
a space-time theory of gravity, got established on a firmer footing than any other theory in physics. Gravitational
waves are just propagating disturbances in the gravitational field of extremely strong sources caused by some catastrophic 
event associated with cosmic bodies, like binary black hole coalescence, or neutron star mergers. As these events happen
very far away in cosmos, hundreds of millions of light years away and the signal strength would be extremely weak, it
requires extraordinary detection and analysis technology to observe an event. Luckily  the joint collaboration  
LIGO-VIRGO, have so far detected two events in September and December of 2015 during their analysis of observations
made with the laser interferometers over the last few observing sessions. The talk will give a brief theoretical 
sketch of the analysis required for describing the waves resulting from mass motion in the realm of general relativity,
and point out, the serious and sincere efforts of the past fifty years that went into the final success. An attempt will
be made to point out the enormous scope that is available for the new generation of students and researchers in pursuing
the topic as a new window for possibly viewing the Universe with the implications for the studies of  Dark matter, Dark
energy and Cosmology as a whole. 
\end{abstract}

With Galileo’s telescope, the view of the Universe expanded rapidly from naked eye observations to cover the celestial beauty of  Jupiter’s moons, Saturn’s rings and outer planets to the expanding Universe of Hubble that included, Galaxies, Clusters and myriads of extra galactic objects. In 1930, came the next revolution of the enigmatic Radio Universe, lead  by Karl Jansky and  G Reber, which led one to very high energy cosmic sources like Quasars, Active Galactic Nucleii, Pulsars and more importantly, the most profound all encompassing   `Cosmic  microwave Background' that indicated the beginning of our Universe. Once the realisation came about  emissions from cosmic sources in two different frequencies, the  optical and the radio, it was a simple task to look for emissions in other frequencies, IR, UV, X-ray sources. Along with came the bonus of  emissions  of  $ \gamma $-rays , which  completed the electromagnetic spectrum of Universe being visible in the entire spectrum from Radio waves to  $ \gamma $-rays. 
While it was known that the emission of radiation from most of  these sources were all due to Electromagnetic processes, it was not clear upto 1960s, the source of energy for emissions from objects like Quasars and AGN s, till Hoyle and Fowler put forward the idea of  Gravitational collapse of massive stars, within the framework of  Einstein’s theory of General Relativity which explained Gravitation as the curvature of  space-time.\\

Of all the discoveries of the human mind, Einstein’s theory of general relativity is considered to be the most beautiful creation. In fact, it is often said that the special relativity, which forms a strong basis of modern physics, along with quantum mechanics, was ripe to be discovered at the turn of the nineteenth century, and if not Einstein, Poincare or Lorentz  would have developed the theory. On the other hand the general theory of relativity, which is the epitome of the world of symmetry, assigning freedom from the confines of coordinate systems (observers) to understand the most important of all the fundamental interactions-Gravity, is completely the work of one individual, arising out of thought experiments instead of laboratory experiments or observations, that preceded all other discoveries in physics.\\

The most important features that lead Einstein from special to general relativity are-\\

\begin{figure}[h]\centering
\includegraphics[width=10cm]{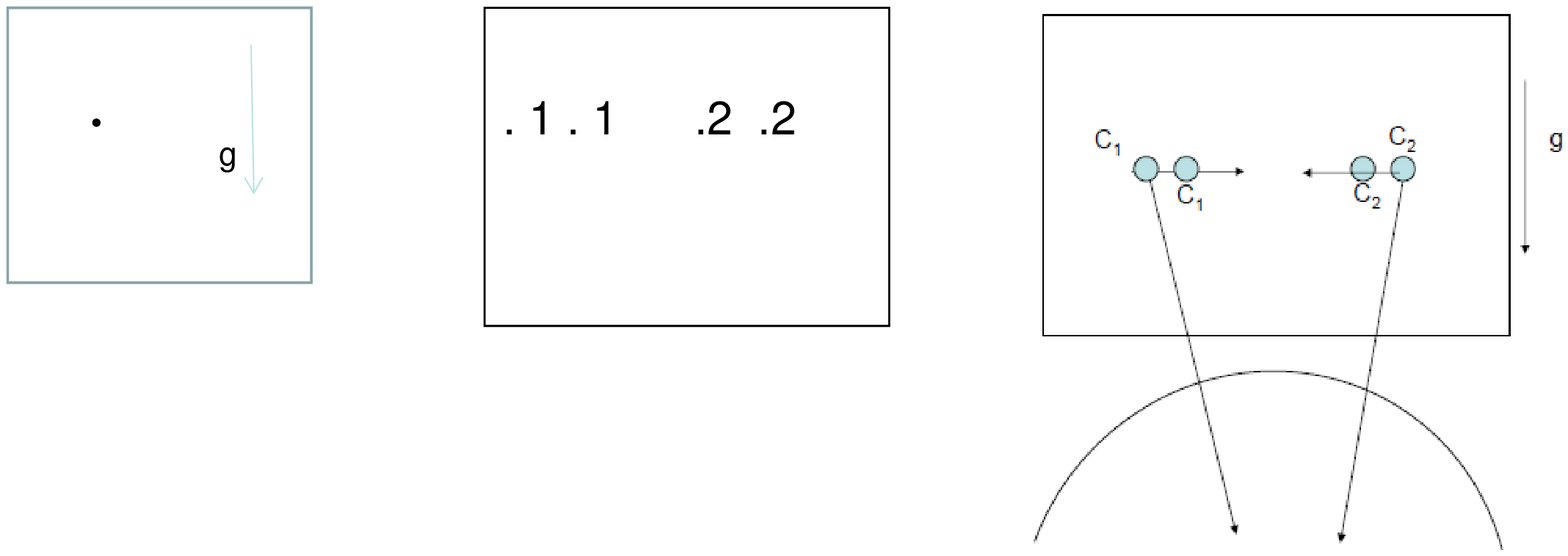}
\caption{Freely falling elevator \cite{arp08}} \label{ffe1}
\end{figure}

1.The equivalence of inertial and gravitational mass of any body   $ M_i = M_g  $ known as the principle of  equivalence (demonstrated by Eotvos in 1889), also guided by his thought experiment of a freely falling lift with an observer inside.\\
 Consider the observer in the elevator  dropping two coins side by side as shown in figure \ref{ffe1}.The observer  inside will see them hung in air($ a = g $) as both the coins and the elevator are falling with the same acceleration. However, if the observer has a very accurate measuring rod, he will find that with time the two coins appear coming closer towards each other. The explanation to this is quite simple. As the earth's gravitational field is radially symmetric the paths of the freely falling coins are along lines converging at the center of the earth. It is thus that they appear to be moving towards each other. This means, though the observer cannot measure the gravitational field inside the elevator, he can measure the variation in the field from point to point, or the \textit{relative acceleration} between the particles.

\begin{figure}[h]\centering
\includegraphics[width=6cm]{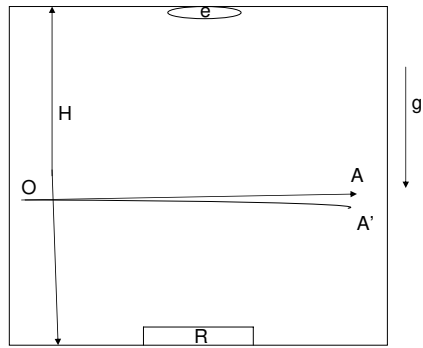}
\caption{light beam in a freely falling elevator \cite{arp08}} \label{ffe2}
\end{figure}

\begin{figure}[h]\centering
\includegraphics[width=8cm]{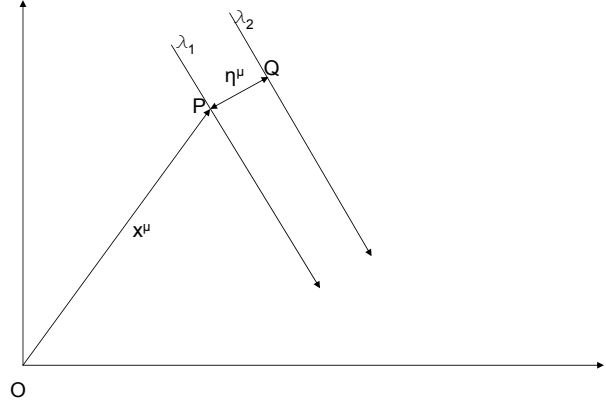}
\caption{two freely falling particles \cite{arp08}} \label{2ffpars}
\end{figure}

\begin{figure}[h]\centering
\includegraphics[width=8cm]{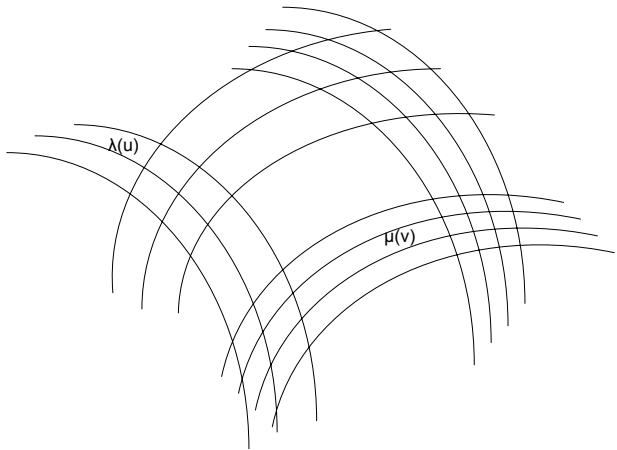}
\caption{congruence of world lines} \label{2sfce}
\end{figure}

\begin{figure}[h]\centering
\includegraphics[width=10cm]{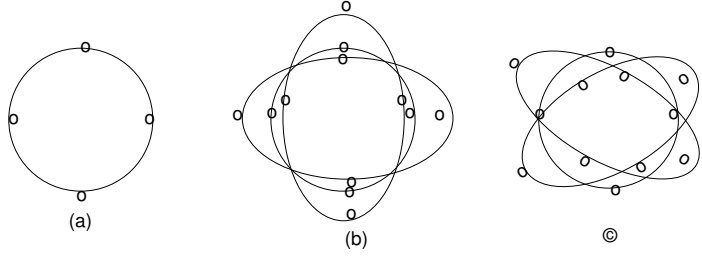}
\caption{gravitational wave passing thrugh a ring of particles. (a) Before, (b)wave with + polarisation, and (c) wave with x polarisation} \label{gwvring}
\end{figure}

 2. effect of gravity on light (figure \ref{ffe2}) which implies that light has a curved path in a gravitational field, as also there appears a frequency shift depending upon the gravitational potential difference.
\begin{equation}\label{freqrel}
  (\nu_o - \nu_e)/ (\nu_e) =  v/c = -gH/c^2,
\end{equation} 
With this Einstein realized that the arena he wanted for general relativity, was the non Euclidean geometry, (non flat),which requirement was satisfied by the Riemannian Geometry, an extension of  Gaussian curved geometry, as suggested by his mathematician friend Marcel Grossmann, represented by the metric $ ds^2 = g_{ij} dx^i dx^j $. As one saw in the case of the elevator, two freely falling particles in a gravitational field have a relative acceleration which in Newtonian terms can be seen as follows. Consider two freely falling particles in a gravitational field with their trajectories labeled  $ \lambda_1 $ and $\lambda_2 $. At time $ t $ let their positions be  P and Q respectively, identified in the associated frame by   P $(x^a)$ and  Q $(x^a + \eta^a)$, all  coordinates being functions of $ t $ [figure \ref{2ffpars}].
 \\
$ \eta^a $ is a small connecting vector between $ \lambda_1 $ and $\lambda_2 $,   $ t $ being the path parameter. One can write, their equations of motion to be
\begin{equation}\label{relacn1}
\ddot x^a  = - (\partial^a \phi)_P = -\delta^{ab}\partial_b (\phi)_P\\ 
\end{equation}
\begin{equation}\label{relacn2}
\ddot x^a + \ddot \eta^a  =  - (\partial^a \phi)_Q = -\delta^{ab}\partial_b (\phi)_Q
\end{equation}
where $ \phi $ is the gravitational potential in which the particles are falling.\\
As $\eta^a $ is very small, one can use Taylor expansion and simplify to get 
\begin{equation*}
\ddot \eta^a = - \eta^k \partial_k \partial^a \phi,
\end{equation*}
which may be written as 
\begin{equation}\label{relacn3}
\ddot\eta + K^a_{\phantom {a} b}\eta^b = 0 ;\quad  K^a_{\phantom {a} b} = \partial^a \partial_b \phi.
\end{equation}
$\ts K{^a_b} $ thus represents the relative acceleration between the two particles as they fall freely in the gravitational field of $ \phi $. \\
Moving on to general relativistic formulation, one can consider a two surface made of a congruence of geodesics and some connecting vectors $ \eta^i $ as shown figure \ref{2sfce}.
 Using the equation for absolute derivatives one can get the geodesic deviation equation \begin{equation}\label{relaccn3}
\frac{\delta^2 \eta^i}{\delta s^2} +  R^i_{\phantom{i} l k j}U^l \eta^j U^k = 0.
\end{equation}   
and then referring it to a local Lorentz frame, one can find the acceleration equation to be
\begin{equation}\label{relaccn4}
\frac{d^2 X^{(a)}}{ds^2}  + K^{(a)} _{(c)} X^{(c)} = 0,  \qquad  K^{(a)} _  {(c)} = \ts R{^{(a)} _{(b)(c)(d)}} u^{(b)} u^{(d)}.
\end{equation}  
\\The field   equations of general relativity are \,[ $ \ts G{_{ij}}\equiv \ts R{_{ij}}-\frac{1}{2} R \ts g{_{ ij}} = \kappa \ts T{_{ij}} $]\, along with the Bianchi identities \quad [$ \ts G{^{ij}_{; j}} = 0 $], \, and consequently the conservation laws \, [ $ \ts T{^{ij}_{; j}} = 0 $]. \\
    If the mass distribution is static, then for some specific symmetrical distributions, a few exact solutions have been obtained as given by the Schwarzschild (spherically symmetric) and the Kerr solutions (axisymmetric) apart from a few cylindrically symmetric solutions, for the cases of uncharged and charged matter that are asymptotically flat. On the other hand, when the matter distribution is nonstatic, then one expects the field surrounding the distribution to vary slowly and the change could propagate all through the space--time as small perturbations of the background field. In such a situation, one could write the general metric solution as, $ g_{ij} = g^b _{ij} + h_{ij}. $  As Einstein had pointed out, things are made much simpler when one assumes, the background metric to be flat, $ g^b _{ij}= \eta_{ij}.$\\
 Considering the field outside the matter distribution, and substituting for the components of the Ricci tensor, the metric $\eta _{ij} + h_{ij} $ and its derivatives in the equations, $ R_{ij} = 0,$ one finds for the perturbations $ h_{ij} $ the set of equations \begin{equation}\label{lineqns1}
  \square h_{ij} + \ts h{^k _k _{, i j}} - \ts h{^k _i _{, j k}} -\ts h{^k _j _{, i k}} = 0,
   \end{equation}
where $ \square $  represents the usual flat space D'Alembertian, with the second and higher powers of $ h_{ij} $ being ignored.\\
It is understood that solutions to this equation cannot be unique, as one can have a general coordinate transformation, and in order to remove this ambiguity, one can choose a particular gauge, and one often chooses the so--called, harmonic or Lorentz (also called deDonder) gauge, as given by $ g^{ij} \ts \Gamma{^k _{ij}} = 0, $ which in terms of $ h $ yield the relations,\begin{equation}\label{gaugecon1}
 \bar h^j _{i,j} = 0,  \quad  \bar h^j _i = h^j _i - \frac{1}{2}\delta^j _i h^k _k.
 \end{equation}
 With this choice of gauge, the equations reduce to the simple flat space wave equation for the tensor potential, $ h_{ij} $,
  \begin{equation}\label{wveqn1}
  \square h_{ij} = 0.
\end{equation}   
for which one can write the general solution as a superposition of plane monochromatic waves,\begin{equation}
h_{ij} = A_{ij} e^{i k_l x^l} + A^*_{ij} e^{- i k_l x^l},
\end{equation}   
with $ A $ and $ A^* $ representing the complex amplitudes and $k^l $ the wave covector, satisfying the orthogonality relation, $ \eta_{ij} k^i k^j = 0. $\\
The gauge condition yields four constraints on the ten complex amplitudes, given by the  relation \begin{equation}\label{gaugecon2}
 A_{ij} k^j = \frac{1}{2} A^j _j k_i.
 \end{equation}
However, as the coordinate freedom is still left within the gauge as specified by \begin{align}
\square \xi_i = 0, \quad A' _{ij} = A_{ij} + k_i \xi \hat \xi_j + k_j \xi \hat \xi_i,\\
\xi^k (x) = i[\hat\xi^k e^{ (i k_l x^l)} + \hat \xi^{*k} e^{ (i k_l x^l)}], 
\end{align} 
where $ \hat \xi $ are constants by choosing them appropriately, one can make four of the $ A_{ij} $s zero. In order to remove this freedom, one needs four additional constraints, which is achieved by choosing a globally defined time like vector field $ u^i $ such that, $ A^i _j u^j = 0, \,  A ^i _i = 0  $. Thus there are eight constraints on the ten complex amplitudes as given by \begin{equation}
A_{i j} u^j = 0, \quad A_{i j} k^j = 0, \quad   A ^i _i = 0, 
\end{equation}
indicating that the $ A_{ij}$s are transverse and traceless. Such a choice of gauge is known as T--T gauge or transverse traceless gauge.\\ 
In terms of the metric potentials, the choice of T--T gauge yields, \begin{equation}
h_{i0} = 0, \quad  \ts h{_a^j_{, j}} = 0, \quad h^i _i = 0.      
\end{equation}
As there are only two degrees of freedom associated with the waves, it implies physically that there are only two degrees of polarisation associated with these waves.Thus, for a plane gravitational wave propagating along the z-direction, in a Cartesian system, the solution may be written explicitly as \begin{align}\label{pgwamp}
 \ts h{^ {TT} _{XX}} = -\ts h{^ {TT} _{YY}} = \sR \{a_+ \,e^{[-i\omega (t-z)]}\} \nonumber\\
 \ts h{^ {TT} _{XY}} = \ts h{^ {TT} _{YX}} = \sR \{a_\times  e^{[-i\omega (t-z)]}\} 
 \end{align}
with $ a_+ = A_{11} = - A_{22} $   and  $ a_\times = A_{12} = A_{21}, $
denoting the two independent states of polarisation.\\
In the case of a monochromatic plane gravitational wave, propagating along the z-direction, the space-time metric is given by \begin{equation}\label{relaccn2}
ds^2 = dt^2 - (1- h_{XX}) dx^2 - (1 - h_{YY}) dy^2 + 2 h_{XY} dx dy  - dz^2 
\end{equation}
and the only components of the curvature tensor that are nonzero are \begin{align}\label{rhijk}
(i)\quad   &\,\ts R{^x _{0 x 0}} = -\frac{1}{2} \ts h{^{TT} _{XX,00}}, \\
(ii)\quad  &\,\ts R{^y _{0 y 0}} = \frac{1}{2} \ts h{^{TT} _{YY,00}}, \\
(iii)\quad &\, \ts R{^x _{0 y 0}} = \,\ts R{^y _{0 x 0}} = -\frac{1}{2}\ts h{^{TT} _{XY,00}}. 
 \end{align}
 Choosing a comoving frame $ u^i = (1, 0, 0, 0) $ and the deviation vector $ \eta^i = (0, \varepsilon, 0, 0),$ equation \eqref{relaccn3} yields\begin{align}\label{eometa1}
 (i)\quad  & \frac{\partial^2\eta^x}{\partial t^2} = \frac{1}{2}\ts h{^{TT} _{XX,00}}\;\varepsilon,\\
 (ii)\quad  & \frac{\partial^2\eta^y}{\partial t^2} = \frac{1}{2}\ts h{^{TT} _{XY,00}}\;\varepsilon.
 \end{align}
On the other hand if the deviation vector $ \eta^i = (0, 0, \varepsilon, 0) $ then the equations are,
\begin{align}\label{eometa2}
(iii)\quad & \frac{\partial^2\eta^x}{\partial t^2} = \frac{1}{2}            \ts h{^{TT} _{XY,00}}\;\varepsilon,\\
(iv)\quad  & \frac{\partial^2\eta^y}{\partial t^2} = -\frac{1}{2}           \ts h{^{TT} _{XX,00}}\;\varepsilon.
 \end{align}
 It is clear from these four equations that the passing wave induces oscillations of the particles in the ring depending upon the nonzero components of the tensor $ h_{ij}. $\\
  If  $ h_{XY} = 0, $ and $ h_{XX} = - h_{YY} \neq 0, $ then the ring of particles oscillates as shown in figure  \ref{gwvring}(b), along the X and Y directions, with $ h_{XX} $ changing sign.
 On the other hand if the wave is such that, $ h_{XX} , h_{YY} $ are zero but $ h_{XY} \neq 0, $ then the particles oscillates as shown in figure  \ref{gwvring}(c).\\
\\\textbf{Do these waves carry energy and angular momentum?}\\

As has been pointed out in references \cite{MTW}, \cite {weinberg72}, \cite{ll75},  the gravitational field energy cannot be localised and thus it is difficult to separate the source energy and the field energy from the total energy momentum tensor $ T^{ij} $ that appears in the field equations. However, in the case of waves as described here, one has an advantage that in the linearised theory, one can  still construct a pseudotensor\index{pseudotensor} that characterises the energy momentum for gravitational waves.\\
  Writing the general energy momentum conservation law coming from the field equations, $ \ts T{_i ^j _{;j}} = 0, $ as \cite{ll75}
 \begin{equation}\label{6conslaw}
 \frac{1}{\sqrt{-g}} [\frac{\partial (T_i ^j \sqrt{-g}}{\partial x^j}] - \frac{1}{2} \frac{\partial g_{jk}}{\partial x^i} T^{jk} = 0,
 \end{equation}
one can see that it gives the simple conservation law for the source when the potentials $ g_{ij} $s are constants. Rewriting it as \begin{equation}
\frac{\partial}{\partial x^j}[(-g) \, ( T_i ^j + t_i ^j)] = 0,
\end{equation}
one can see that the total energy momentum has been separated into a part representing the source energy and the remaining the field energy $ t_i ^j $  called the Landau--Lifshitz pseudotensor, obtained from a  super potential, $ \Psi^{i k l} $, defined through  the equation \begin{equation}\label{llpsudotens}
(-g) [( R^{ik} - \frac{1}{2} R g^{ik}) + t^{ik}] = \ts \Psi{^{i k l} _{, l}}.
\end{equation}
 The L--L super potential, when expressed in terms of the metric and its derivatives, is given by \cite{anderson72} \begin{equation}\label{psidef}
 \ts \Psi{^{i k l}} = \sqrt{-g} \,\delta^i _p \,\{g (g^{kp} g^{lm} - g^{km} g^{lp})\}_ {, m}. 
\end{equation} 
With this definition, one then calls the total energy momentum, $ (-g) \,(T^{ik} + t^{ik})$  the `effective energy momentum', of the space-time governed by the chosen metric, that satisfies the usual divergence--free relation, $ (T^{ij} + t^{ij})_{,j} = 0, $  such that one can use the volume integral and recover the effective energy. \\
 In the case of linearised gravity, with the perturbations defined over a flat background metric ( $ g_{ij} = \eta_{ij} + h_{ij} $), as shown in \cite{MTW}, for the short wave approximation,  defined by ($ \lambda / \sR \ll 1,  a \ll 1 $), the effective stress tensor averaged over several wavelengths is given by
  \begin{equation}
 t_{ij} = \frac{1}{8\pi}\{< R_{ij}(h^2)>- \frac{1}{2} \ts g{^B_{ij}}<R (h^2)> \}, 
\end{equation}
which, for the flat background, yields, in the TT gauge the expression, \begin{equation}
<t_{ij}> = \frac{1}{32 \pi} < \ts h{^{kl} _ {,i}} \,  \ts h {_{kl , j}}>.
\end{equation}
This is also commonly referred to as Issacson stress--energy tensor for gravitational waves, \cite{schutz}, \cite{isac68}, when the averages are taken over one period of oscillation in time and spatial regions of the size of a wavelength of distance in all directions.\\
 Going back to the field equations one can see that on using the harmonic gauge   $ \bar h^j _{i , j} = 0 $, the equations reduce to 
\begin{equation}\label{6fldeqns2}
\square \,\bar h_{ij} = -2 \kappa \tau_{ij},   
\end{equation}
whose integrability  requires $ (\tau_i ^j)_,j = 0. $\\

One can  write the solution of \eqref{6fldeqns2} in terms of a retarded Green's function, \cite{demansk85}which, after integration with respect to t', yields,\begin{equation}\label{6soln2}
\bar h_{ij} = 4 \int\{[\frac{\tau_{ij} (x',\,t -\vert x - x' \vert )]}{\vert x - x'\vert}\} \, d^3 x'.
\end{equation}
As $ \tau_{ij} $ satisfies the conservation law, $ \tau^{ij} _{, j} = 0, $
one can write this as \begin{align}\label{6conslaw}
(i)\quad  \ts \tau{^{ab} _{,b}} + \ts \tau{^{a0} _{,0}} = 0 \nonumber \\ 
(ii)\quad  \ts \tau{^{0b} _{,b}} + \ts \tau{^{00} _{,0}} = 0.
\end{align}
indices a,b taking values 1,2,3.
Taking the appropriate moments of these equations and simplifying one gets finally\begin{equation}\label{2mom}
\int \tau^{ab} dV = \frac{1}{2} \frac{\partial^2}{\partial t^2}\int  \rho(r',t) x^a x^b dV = \frac{1}{2} \ddot I^{ab},
\end{equation}
where $ I^{ab} $ is the second moment of the mass distribution  at the source related to the moment of inertia tensor \cite{MTW}, 
\begin{equation}\label{mominr}
\sI^{ab} = \int \rho(r^2 \delta^{ab}- x^a x^b) dV = (\delta^{ab}I^c _c - I ^{ab}),
\end{equation}
and to the quadrupole moment $ Q^{ab} $ \cite{ll75}
\begin{equation}
Q^{ab} = \int \rho(x,t) (3 x^a x^b - r^2 \delta^{ab}) dV = (3 I^{ab} - \delta^{ab} I^c _c).
\end{equation}
With this one can finally write down the approximate solution for \eqref{6fldeqns2}, \begin{equation}\label{quadform}
\bar h_{ij} = \frac{-2\Omega^2}{r} I_{ij} e^{[i\Omega (r-t)]},
\end{equation}
$\Omega $ being the frequency. 
Equation \eqref{quadform} is the well known `quadrupole formula' for gravitational radiation. \\
The individual components of the metric tensor $ h_{ij} $ for a plane gravitational wave in T--T gauge, moving along the z-direction, are now given by \begin{align}
h_{Zi} = 0,(i= 0,1,2,3);\quad    h_{XY} = -\frac{2\Omega^2}{3r} Q_{XY} e^{[i\Omega(r-t)]}\nonumber\\ h_{XX} = - h_{YY} = -\frac{\Omega^2}{3r} (Q_{XX} - Q_{YY}) e^{[i\Omega(r-t)]},
\end{align}
and the energy flux carried along the direction of propagation is \begin{equation}
t^{z0} = (\frac{G}{36 \pi r^2 c^5}) [(\frac{\dddot Q_{XX} - \dddot Q_{YY}}{2})^2 + (\dddot Q_{XY})^2].
\end{equation} 
  In order to express the energy and angular momentum carried by the waves, in an invariant form, one can use the 3-dim. symmetric, unit polarisation tensor $ e_{ab} $ \cite{ll75}, which determines the nonzero components of the metric tensor $ h_{ab} $ in the appropriate gauge  ($ h_{0a} = h_{a0} = h = 0) $ and satisfies the relations 
  \begin{equation}\label{poltens1}
  e_{0a} = 0,  \quad  e_{ab} n^b = 0,  \quad  e_{ab} e^{ab} = 1, 
  \end{equation}
$ n^a $ being the unit  vector along the direction of wave propagation.\\ 

 The intensity of radiation of a given polarisation into a given solid angle $ d\Sigma $  is then \begin{equation}\label{intensity1}
 dI = \frac{1}{72 \pi} (\dddot Q_{ab} e^{ab})^2 d\Sigma,
 \end{equation}
which depends implicitly on the direction $n$, because of the condition of transversality. Summing over all polarisations then gives the total angular distributions,\begin{equation}\label{intensity2}
dI = (\frac{G}{36 \pi c^5})[ \frac{1}{4}(\dddot Q_{ab} n^a n^b)^2 +\frac{1}{2}\dddot Q_{ab} ^2 - \dddot Q_{ab} \dddot Q^a _c n^b n^c ].
\end{equation}
The energy loss of the system per unit time can be found by averaging $\frac{dI}{d\Sigma}$ over all directions and multiplying by $ 4\pi $ as  given by 
 \begin{align}\label{eneangmm}
&\frac{dE}{dt} = -(\frac{G}{45 c^5}) <\dddot Q_{ab} \dddot Q^{ab}> \\
&\frac{dJ_k}{dt} = -(\frac{2G}{45 c^2} )\varepsilon_{klm}\, <\dddot Q^{la} \dddot Q_a ^m> .
\end{align}
 In the quadrupole approximation, which also happens to be the lowest order post--Newtonian approximation, one can express the amplitude, frequency, and luminosity of the emitted radiation \cite{sslr12}, which depend only on the density $ \rho $ and velocity fields of the Newtonian system as given by\\ 
(a) the amplitude in the Lorentz gauge,\begin{equation}
 h_{ab} = \frac{2}{r}\frac{d^2 Q_{ab}}{dt^2}, \quad Q^{ab} = \int \rho x^a x^b \,d^3x,
 \end{equation}
(b) the frequency, \begin{equation}\label{gwlumin}
 f_0 = \omega_0 / 2 \pi = \sqrt{G \bar \rho /4\pi},
\end{equation}
where $ \bar\rho $ is the mean density of mass--energy in the source.\\ (c) the luminosity expressed in terms of the local stress--energy in the T--T gauge is given by 
\begin{equation}
L_{gw} =\frac{1}{5}(\Sigma_{j,k} (\dddot Q_{jk})^2 - \frac{1}{3} (\dddot Q)^2,
\end{equation}
where Q is the trace of $ Q_{jk},$ an equation which may also be used to estimate the back reaction on a system emitting gravitational radiation\cite{sslr12}.\\

\textbf{Detection of gravitational waves}\\

Late last year (2015), the world celebrated the one hundredth anniversary of general relativity and Einstein's prediction of gravitational waves by detecting through LIGO (Laser Interferometric Gravitational wave Observatory) the first signal of the waves arriving on earth, produced far away in the cosmos, by coalescence of two medium--sized black holes .\cite{abbott16}.\\
 As one can see from the expression for the energy carried by the wave \eqref{eneangmm}, its strength is of  order $ c^{-5} $, and thus would require an extremely sensitive set of apparatus and very sophisticated methods of data analysis to detect signals of such low strength and to separate them from all other forms of noise.\\
 
  The experimental search for gravitational waves from cosmic sources started with J. Weber's pioneering idea of using a resonant bar detector \cite{weber60}, which was essentially  a suspended homogeneous  metal bar, on which an impinging gravitational wave would excite mechanical vibrations that could be transferred to electromagnetic signals by piezoelectric transducers which can be amplified and recorded. The excitation is mainly due to the relative acceleration between the particles of the bar caused by the passing wave. When two such antennas separated by a large distance (in the case of Weber, the bars set up were in Maryland, Virginia and Argonne National lab in Chicago) record similar signals coincidentally, it was assumed that the disturbance was caused by a cosmic source far away from the earth and attributed to gravitational wave.
\begin{figure}[h]\centering
\includegraphics[width=10cm]{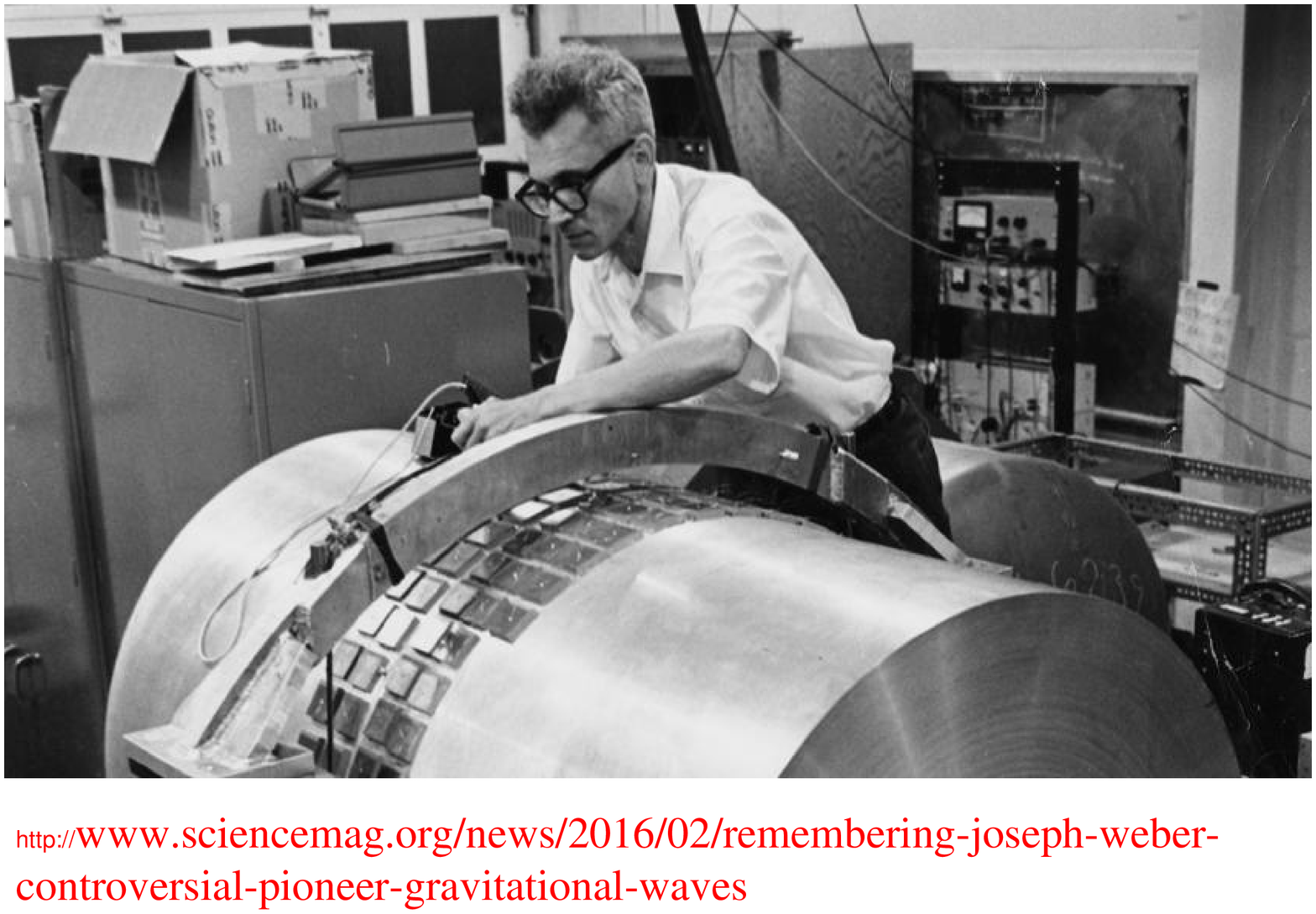}
\caption{Joe Weber and his antenna} \label{jweber}
\end{figure}  
   (Although Weber announced the recording of such signals in 1969  claiming the detection of gravitational waves, it was very soon found to be not correct as no other experimental group, even with increased sensitivity systems could find any coincident signal.) \\
  
   Though there have been continuous efforts to improve the sensitivities of the bar mode detectors, the attention of the experimental community turned towards the beam mode detectors,\index{beam detectors} where one uses laser interferometry, consisting of four masses hung from vibration--free support systems with their separation being monitored by a highly sophisticated optical system. 
\begin{figure}[h]\centering
\includegraphics[width=10cm]{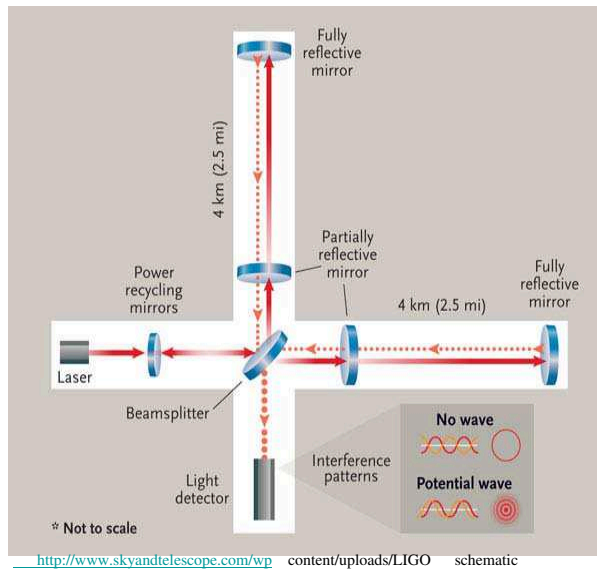}
\caption{LIGO interferometer} \label{ligo}
\end{figure} \\   
   The four masses (mirrors) are placed at the ends of two orthogonal arms such that two are closer to  each other with the other two at the far ends of the arms, and the arm's lengths being almost equal ($ L_1 \simeq L_2 = L, $ such that) the change  ($ \bigtriangleup L(t), $ is directly proportional to the output of the interferometer (photodiode). When a gravitational wave passes through such a system, having frequency higher than the pendulum's natural frequency of $ \sim 1 Hz $, the acceleration induced by the wave pushes the masses (as though they are freely falling)  which causes the arm length difference $ \bigtriangleup L = L_1 -L_2 $ to change. Depending upon the polarisation of the impinging wave, ($ h_+ or \,h_\times $)  the interferometer's output would be a linear combination of the two wave fields \cite{thorne97}
    \begin{equation}\label{deltal}
   \frac{\bigtriangleup L (t)}{L} = F_+ h_+ (t) + F_\times h_\times (t)\equiv h(t)
 \end{equation}
where $ F_+, F_\times $  are of order unity having quadrupolar  dependence upon the direction and orientation to the source \cite{thorne87}. The $ h(t)$ in \eqref{deltal} is called the strain of the gravitational wave and the time evolution of $ h(t), h_+(t), h_\times (t) $  as waveforms. \\
\begin{figure}[h]\centering
\includegraphics[width=10cm]{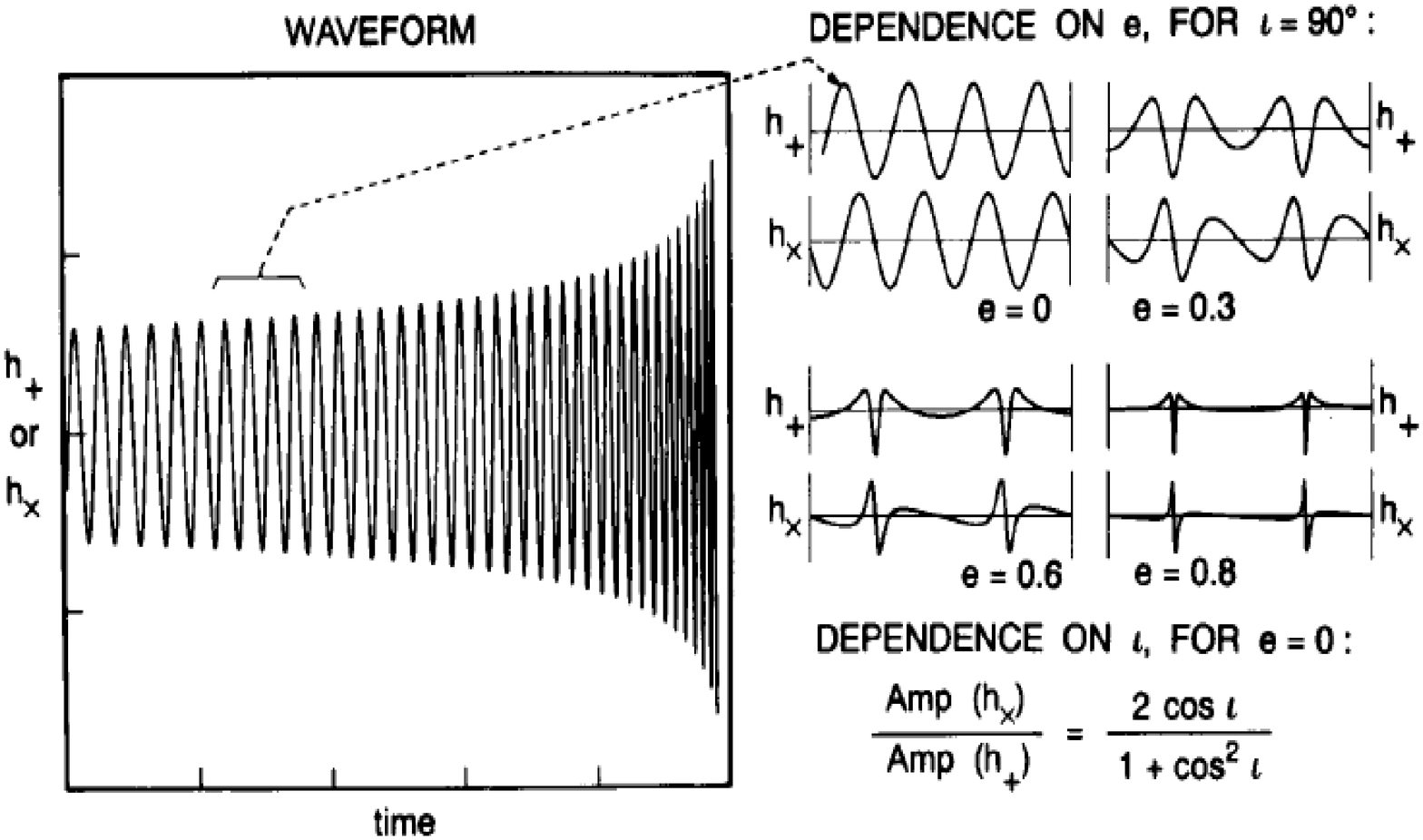}
\caption{Typical waveforms from The inspiral of a compact binary computed using Newtonian gravity for the orbital evolution and the qudrupole moment approximation for the wave  generation \cite{abramwvfm}} \label{6wvform}
\end{figure} \\
A typical waveform arising out of inspiraling compact binary system\index{inspiraling compact binary} appears as shown in figure \ref{6wvform}, which has been computed using Newtonian gravity for the orbital evolution and the quadrupole-moment approximation for wave generation \cite{abramwvfm}. As the inspiralling binaries get closer, one finds  increasing amplitude and the upward sweeping frequency (often referrred to as chirp) of the   waveform, with the amplitude ratio for the two polarisations going as \begin{equation}
\frac{amp\, h_+}{amp\, h_\times} = \frac{2 \cos i}{1 + \cos^2 i},
\end{equation}
$ i $ being the inclination of the orbit to observer's line of sight, and the orbital eccentricity determining the waves' harmonic content. For simplicity, if the orbit is considered circular, then the rate at which the frequency sweeps  or `chirps'\index{chirp}, $ df/dt $ (also referred to as the number of cycles spent near a given frequency $ n = f^2 (df/dt)^{-1}, $ is determined solely by the binary's chirp mass in terms of the masses of the binary components, $ M_c \equiv \frac{(M_1 M_2)^{3/5}} {(M_1 + M_2)^{1/5}} $. Thus, the amplitudes of the two waveforms ($ h_+, h_\times) $ are determined by  the chirp mass, distance to the source, and the orbital inclination. With the preliminary information coming from the qudrupolar (near Newtonian) formula, the general relativistic effects add further information, through the waveform modulation coming from the rate of frequency sweep, depending upon the binary's dimensionless ratio, $ \eta = \mu /M,$ with $ \mu = M_1 M_2 / (M_1 + M_2) $ the reduced mass and $ M = (M_1 + M_2) $, the total mass, as well as on the spins of the two bodies. Two of the important effects worth noting are (i)  the back scattering of waves due to the curvature of the binary space--time \cite{vish70}, producing tails that act back on the binary modifying  the inspiral rate that can be measured and (ii) the Lense--Thirring drag arising from the inclinations of the spin axes of the components with respect to the binary's orbital plane, causing the orbit to precess, which, in turn can modulate the wave forms. In order to incorporate these relativistic modulations of the basic wave forms, while detecting, one uses a technique called the matched filter,\index{matched filter} where the incoming signals are matched to already prepared theoretical templates \index{template}with several different combinations of parameters, and the best matched template will give the details of the wave form \cite{thorne97}.\\

     As Blanchet points out \cite{blanchlr14}, the basic problem that one faces in relating the amplitude $h_{ab}$ seen in the wave zone with the source material stress energy, $ T_{ij} $, is due to the approximation methods in general relativity. While the post--Newtonian methods may appear satisfactory in the weak field limit (valid only in the near zone), its inadequacy appears while trying to include the boundary conditions at infinity, which affects the proper determination of the radiation reaction force. While the post-Minkowskian approximation appears valid all over the space--time as long as the source is weakly gravitating, it faces hurdles while treating the multipole approximation outside the source with respect to the far zone expansion. \\
     
  In the early 1970s, while several groups were still trying to check Weber's claim of the detection of gravitational waves, an altogether different set of observations confirmed the existence of gravitational waves indirectly. Hulse and Taylor, during a routine search for pulsars, from the Arceibo Observatory had recorded several new pulsars, amongst which was the discovery of the first binary pulsar PSR 1913 + 16, which was identified as a set of two neutron stars with almost equal masses, ($ M_p = 1.39\pm 0.15 M_\odot,\, M_c = 1.44\pm 0.15 M_\odot), $ moving on a fairly eccentric orbit ($ e = 0.617155\pm 0.000007 $), quite close to each other having the projected semi--major axis $ a \sin i \sim 7\cdot 10^{10} cm$ \cite{ht75}. Continuous monitoring of the binary pulsar over the next few years, yielded a much better evaluation of the orbital parameters \cite{taylor79}, which clearly revealed the binary pulsar system to be the best laboratory for testing general relativity. As summarised by Weisberg and Taylor (2005) the measured orbital parameters over the period, 1981-- 2003, are as listed in the table below \cite{wt05}\\

\begin{table}[ht]
\caption{Measured Orbital Parameters for B 1913 + 16 System}
\centering
\begin{tabular}{   c                          c }
\hline 
fitted parameter                      &         value     \\ 
\hline
\\$ a_P \sin i $(s)                     &      2.3417725 (8)\\

\\$ \omega_0 $                          &    292.54487 (8)\\

\\$ e $                                 &      0.6171338   \\     
\\$ \langle \dot\omega\rangle $(deg/yr) &      4.226595 (5)\\
\\$ T_0$                                &  52144.90097844(5)\\
\\$ \gamma $(s)                         &      0.0042919 (8)\\
\\$ P_b $                               &      0.322997448930(4)\\

\\$  \dot P_b(10^{-12}s/s)$             &     -2.4184(9) \\

\\ 
\hline
\end{tabular}
\label{table 1}
\end{table}  
  
While the first five parameters of the table are derivable purely from  non--relativistic analysis, the next three, the mean rate of advance of periastron $\langle \dot \omega^i\rangle $, gravitational redshift and time-dilation parameter $ \gamma $ and the orbital period derivative $ \dot P_b $ come only from general relativistic corrections.
   
  One of the most important results pointed out was the fact that the orbital period of the system was changing as given by $ \dot P_b $, which can happen only with the loss of the orbital energy bringing the two components closer. Taylor et al. found  the secular decrease of the orbital period to be consistent with loss of energy through emission of gravitational radiation as predicted by general relativity \cite{tw82},  which is calculated on the basis of suggestion from Wagoner\cite{wago75}, and Esposito \& Harrison\cite{esphar75}, using  the analysis of Peters and Mathews \cite{pm63}, as given by \begin{align}\label{pbdot}
\dot P_b = -\frac{192 \pi G^{5/3}}{5 c^5}({\frac{P_b}{2\pi}})^{-5/3} ({1-e^2})^{-7/2}\cdot (1+\frac{73}{24}e^2 + \frac{37}{96})\nonumber\\ e^4 [m_P m_c / (m_P + m_c)^{-1/3}].
\end{align}
 As the relativistic variables $ \langle \dot \omega\rangle $ and $ \gamma $, both measurable quantities, depend upon the masses of the binary components as given by \begin{equation}
\langle \dot \omega\rangle = 3 G^{2/3} c^{-2} (P_b /2\pi)^{-5/3} (1-e^2)^{-1} (m_p + m_c)^{2/3} 
 \end{equation}
and \begin{equation}
\gamma =  G^{2/3} c^{-2} e (P_b /2\pi)^{1/3} m_c (m_P + e m_c) (m_p + m_c)^{-4/3}, 
\end{equation}
inserting the measured values and solving for the masses, one finds \begin{equation}
m_P = 1.4408 \pm 0.0003 M_\odot,  ; \quad m_c = 1.3873 \pm 0.0003 M_\odot.
\end{equation}
Using these in the above \eqref{pbdot}, one can get the orbital period decay rate to be, $ (\dot P_b)_{GR} = (-2.40247 \pm 0.00002)\times 10 ^{-12} s/s. $ \\
\begin{figure}[h]\centering
\includegraphics[width=10cm]{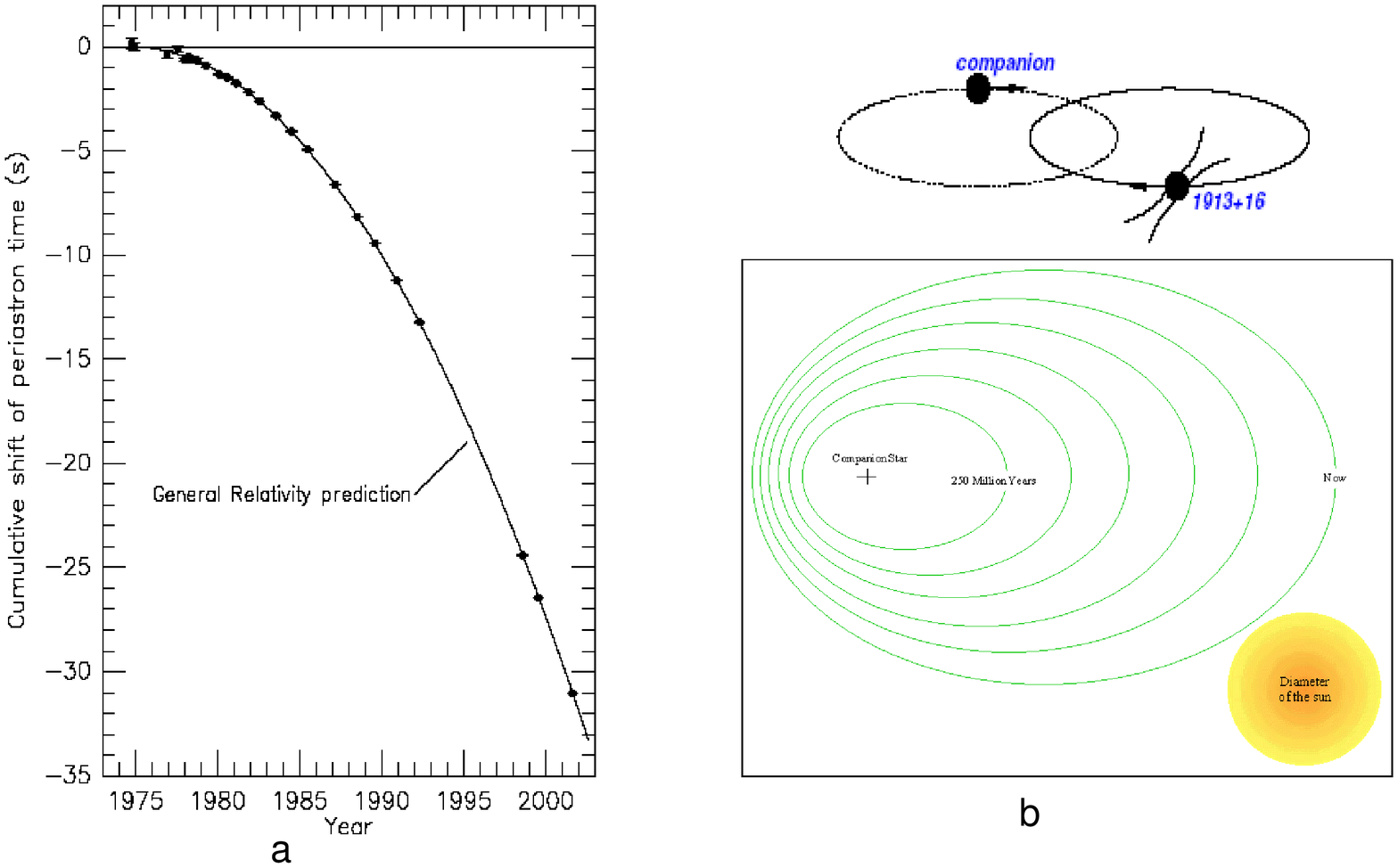}
\caption{(a)Orbital decay of PSR 1913+16 during 1975 to 2003,producing the change in period decay (b) orbit changes leading to coalescence schematic \cite{wt05}.]} \label{orbdecay}
\end{figure}. 

As Damour and Taylor \cite{dt91} argue, there would be some effect on the periods, both theoretical and observational, as a result of galactic acceleration of the system and the motion of the sun, which in fact has several components that add up to $ (\dot P_b / P_b)^{obs} = -86.79 \pm 0.19 (gal) \pm 0.65(obs) 10^{-18}/sec  $, and the corresponding theoretical estimate yields $ (\dot P_b /P_b)^{GR} = -86.0923 \pm 0.0025 (gal) 10^{-18}/sec ,$ yielding the ratio of the observed to the theoretical values of the periods to be\\ 
\begin{equation}
(\dot P_b)^{obs}/(\dot P_b)^{GR} = 1.0081 \pm 0.0022(gal)\pm 0.0076 (obs),
 \end{equation} 
which is an excellent agreement. This orbital decay in period due to gravitational radiation damping should cause a shift in the epoch of periastron as shown in the figure \ref{orbdecay}, where the theoretical curve (solid line) and the observed data points are plotted \cite{wt05}, which shows the remarkable agreement of the data collected over almost thirty years.\\
 According to Blanchet  \cite{blanchlr14}, to observe the final stages of the inspiralling binary coalescence, by the ground based detectors, one requires  very high accuracy templates as predicted by general relativity, and this is achieved by using a higher order post--Newtonian wave generation formalism. This has indeed been achieved to a good degree of applicability, and a host of investigations seem to have demonstrated that the post--Newtonian precision required to successfully implement an optimal filtering technique for the existing detectors (LIGO and VIRGO) to correspond upto 3PN order ($ c^{-6} $) for neutron star binaries, beyond the quadrupole moment.
   (\cite{cutler93}, \cite{fin93}, \cite{tagoshi94}, \cite{poisson97},\cite{krolak95},\cite{damour98}).  
  Whereas these techniques of calculations would suffice to discuss wave emission from binary neutron stars and white dwarfs, they would be found wanting when it comes to the discussion of binary black holes, particularly when one of the components is massive. Modeling the merger of two black holes requires numerical relativity \cite{sslr12}, as calculating the wave forms (templates)requires full solutions of Einstein's equations.\\
 
As reviewed by  Centrella et al \cite{centbaker10} mergers of comparable-mass black-hole binaries are expected to be among the strongest sources of gravitational waves, wherein the final death spiral of a black-hole binary encompasses three stages called inspiral, merger, and ringdown phases. During the inspiral phase, the orbits of the binaries get circularised due to the emission of gravitational waves and further the black holes spiral together in quasi--circular orbits, as the orbitaltime scale would be much shorter than the timescales on which the orbital parameters change. Due to the large separation between the components one can treat them as point particles and thus apply the orbital dynamics as was done for the case of neutron star binaries \cite{peters64}. The wave forms can be calculated using the post--Newtonian equations in terms of   $ v^2/c^2 \sim GM/Rc^2 $,  R being the binary separation \cite{blanchet06}, and one finds that the wave form would have the characteristic of a `chirp', as defined earlier. As the black holes get closer, the weak field limit will not be valid in the merger phase\index{merger phase}, as the strong field dynamical region of general relativity requires the numerical treatment of Einstein's equations 
(a three--dimensional simulation of solutions). At this stage the black holes get close enough to merge and form a single, bigger black hole which could be highly distorted. Finally, this distorted remnant black hole could settle down as a Kerr black hole, after shedding all the nonaxisymmetric modes in the form of gravitational radiation known as `ringdown' phase\index{ringdown}.\\

  The order of magnitude estimates for the amplitudes of the waves emitted at different phases is given by \cite{sslr12}.\\
  
  (a)the inspiral phase;  $ h_b \sim 2 M^2/ r R \simeq ( 2/r) M^{5/3} \Omega^{2/3} $,   (M the mass, R is the orbital radius, r distance to the source, $ \Omega $ the orbital angular frequency), with the luminosity,  $ L_b \sim (4/5 G)(M c/ R)^5 $. As the orbital radius shrinks, the emitted frequency increases towards a chirp, with chirp time for equal mass binary to be $ t_{chirp} = Mv^2 /2 L_b  \sim (5M/8)(M/R)^{-4}$.\\
 
 (b) As the merger stage approaches, with the distance between the components closer to the last stable orbit $(R \sim 6M) $, the frequency reaches the value, $ f_{lso} \sim 220 (20 M_\odot /M)Hz.$ In the case of unequal mass binaries the coalescing  time as measured from the rate of period change, $ \dot P_b = -\frac{192 \pi}{5}(\frac{2\pi\sM}{P_b})^{5/3},$ is $ t_{chirp} =(\frac{5 M}{96 \nu} ( \frac{M}{R})^{-4}$, where M is the total mass of the two components, and $ \sM = \nu^{3/5} M $, the chirp mass, with $ \nu = \mu/M $. one can see from these numbers that, while the binaries with large mass ratios can spend a long time in highly relativistic orbits, those with equal mass are expected to merge  after being in this regime for only a few orbits.\\
 It may be pointed out that the famous binary, Hulse--Taylor pulsar is expected to merge in just  about 300 million years as the orbit is shrinking at the rate of $ \sim 3.1 mm /orbit. $ \\
 
 In the case of massive black hole binaries, as they will be perturbed as they get closer, it is necessary to understand the evolution of black hole perturbations. Vishveswara \cite{vish70} was the first to discuss the consequence of black hole perturbation by the back scattering of the gravitational waves, following an approach initiated by Regge and Wheeler \cite{regwhee57}, for the case of Schwarzschild    blackhole, which was followed up with detailed discussions by Zerrilli, \cite{zer70}, and later for the perturbations of the Kerr metric by Teukolsky \cite{teuk72}. However, the most detailed discussion of the perturbations of blackhole spacetimes was done by Chandrasekhar et al, which can be studied from \cite{chandra}. \\
 
 These perturbed black holes were found to exhibit `quasi--normal modes' of vibration that emit gravitational radiation whose amplitude, frequency, and damping time are characteristic of the black hole`s mass and angular momentum, the only two features of a Kerr black hole. The effective amplitude of the waves is of the form $ h_{eff} \sim \frac{4\alpha \nu M}{\pi r}$, which, for a pair of 10$ M_\odot $ black holes, at a distance of about 200 Mpc, turns out to be $ (\frac{\nu}{0.25})(\frac{M}{20 M_\odot})( \frac{200 Mpc}{r}) 10^{-21}$, and for super massive black holes at cosmological distance is $ 3 \times 10^{-17} (\frac{\nu}{0.25})(\frac{M}{2\times 10^6 M_\odot})( \frac{6.5 Gpc}{r}) $\cite{sslr12}. As the equations of general relativity are a set of coupled nonlinear, second order partial differential equations, the details of the dynamics of the merger of black holes are not accessible for analytic treatment and one resorts to numerical approach.\\
 
     As pointed out by several reviewers, Hahn and Lindquist  \cite{hanlin64} seem to be the first in 1964, to have tried the simulation of the dynamics of, head-on collision of two equal mass black holes, using a two dimensional axisymmetric approach, which they found was not being accurate after 50 time steps. Almost after a decade, Smarr et al. reconsidered the problem, using the ADM formalism (canonical 3+1 formalism \cite{adm62}) with improved coordinate conditions, which led them in spite of the difficulties of instabilities, and large number errors, to some information about the spectrum and total energy of the gravitational waves emitted in the zero frequency limit \cite{smarr77}.\\

  However, the necessity to use numerical methods and computer simulation gained importance with the attempts to detect gravitational waves in the beam detector--like LIGO, in the nineties, as they are sensitive only at the frequencies emitted by black hole mergers. As the signal--to--noise ratios of ground based detectors are fairly modest, constructing templates to pattern the wave forms for this device was very important for data analysis which required numerical simulations. This activated several groups of numerical relativists trying to develop three dimensional codes for relativistic hydrodynamics using super computers which became important \cite{centbaker10},\cite{cghs15}. The successful application of numerical methods and simulations during the period 1990 to 2006, with the revolutionary idea of Pretorius \cite{pretorius05}, advanced the developments in numerical relativity as applied to the detection of gravitational waves immensely, resulting in the final detection of gravitational waves by LIGO/VIRGO collaboration in 2016.\\  
 \begin{figure}[h]\centering 
\includegraphics[width=10cm]{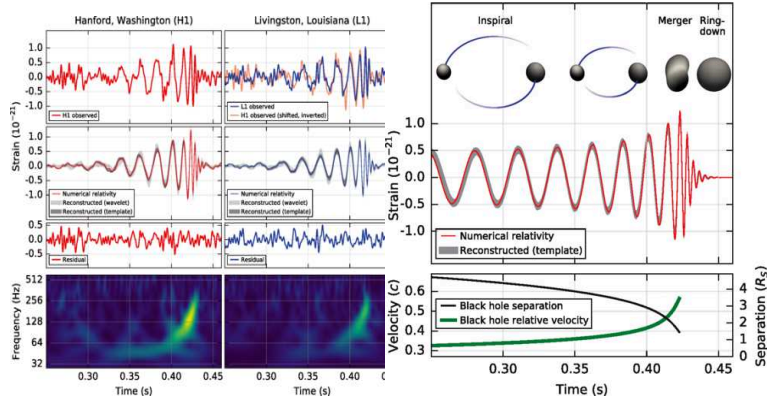}
\caption{The gravitational-wave event GW150914 observed by the LIGO Hanford (H1, left column panels) and Livingston (L1, right column panels) detectors. Times are shown relative to September 14, 2015 at 09:50:45 UTC. For visualization, all time series are filtered with a 35–350 Hz bandpass filter to suppress large fluctuations outside the detectors’ most sensitive frequency band, and band-reject filters to remove the strong instrumental spectral lines seen in the Fig. 3 spectra\cite{abbott16}} \label{grwvdet}
\end{figure}

It is well known that the electromagnetic waves, passing through the intervening matter between the source and the observer, do undergo some changes like frequency modulation, Landau damping etc.,which in a way gives one the information about the medium through which the waves are passing. Can there be any similar effect on gravitational waves passing through matter? Further the analysis above is restricted to purely linearised solution of Einstein’s equations. As the intervening space is not empty it is important to consider the  space-time perturbations for non-flat metric. This question was considered by Ehlers et al, a short discussion of which follows.\\

 \textbf{Propagation of gravitational waves through matter}\\
  
Starting with the perturbations on a non flat background $ g_{ij}= (g_{ij})_B + \hat g_{ij}  $ one can set up the equations for perturbations, for the case of perfect fluid distributionas given by\cite{ep96}
\begin{align}\label{equnspertbs}
\ts P{^{ij} _ {ab}} \hat g_{ij} \equiv &[(2 h^i _{(a} h^c _{b)}\nabla^j \nabla^c - h^i _a h^j _b \nabla^2 - g^{ij} h^c _a h^d _b \nabla_c \nabla_d)  +\alpha h_{ab}(g^{ij} \nabla_u ^2 +\nonumber\\& 2 \nabla^{(i} u^{j)}\nabla_u + 4(\nabla_d u^i \nabla^{[d} u^{j)})) - (\rho - p+ 2\Lambda) h^i _a h^j _b ] (\hat g_{ij}) = 0.       
\end{align}
The characteristic equation, given by the zeroth order equation of the above hirarchy,gives the dispersion relation which is expressed as
\begin{equation}
(g^{ab}l_a l_b)^2 [(u^a u^b - c_s ^2 h^{ab})l_a l_b ] (u^a l_a)^6 = 0.
\end{equation}
shows that there exist three modes\\

(i)  \quad the \textit{gravitational wave} mode, given by the Hamiltonian $ H = \frac{1}{2} g^{ab} l_a l_b$  propagating along the null geodesics having the tangent vector $ T^a = l^a $,\\
 
 (ii) \quad the \textit{sound wave} mode given by $ H = \frac{1}{2} [(c_s ^2 h^{ab}-u^a u^b)] $,\; propagating along the sound rays, with tangent  $ T^a = \omega (\frac{c_s k^a}{k} + u^a)$ , and \\
 
 (iii)\quad the \textit{matter} mode given by $ H = u^a l_a$ , moving along matter rays, $ T^a = u^a $, 
 From the general formalism,that can be referred to in \cite{ep96},one sees that while the zeroth order gives the dispersion relation,the first order gives the transport equation from which one can set for the primary amplitudes the set of ordinary differential equations,
  \begin{equation}\label{trangrwv}
(\nabla_l + \frac{\theta}{2})\begin{pmatrix}
a^{(0)} _+ \\
a^{(0)} _\times
\end{pmatrix} =0.
 \end{equation}
 This implies that the change of the complex vector $ (a_+, a_\times) $ along a ray consists of a rescaling by a positive factor proportional to the square root of the cross--sectional area of a small bundle of rays, just as in the case of gravitational waves in \textit{vacua}. The transport preserves linear, circular, elliptic polarisation, helicity and ellipticity. Further it also implies that the Issacson stress tensor (defined in vacuum)\begin{equation}
 \hat T ^{ab} = \frac{1}{4\pi}(\vert a_+\vert ^2 + \vert a_\times\vert^2) l^a l^b,
 \end{equation}
 which represents the effective energy momentum tensor of the wave, is conserved, $ \nabla_a \hat T^{ab} = 0. $ \\
 The transport equation for the first order primary amplitudes is then given by,\begin{align}\label{transp1}
(\nabla_l + \theta /2) (a_+ ^{(1)}) = &\frac{1}{2}(\rho - p + 2\Lambda)((a_+ ^{(0)}) + \nonumber\\ &+ \frac{1}{2} e^{ij}\{[2 \nabla^ c\nabla_i +\nabla^i \nabla_c)\delta^d _j - \delta^c _i \delta^d _j\nabla^2 - h^{cd}\nabla_j \nabla_i] (v_{0 cd})\}
\end{align}
and a similar one for $ a_\times $.\\

Using the definitions of the curvature tensor, $ \ts R{^h _{ijk}} $, and the Weyl tensor, $ \ts C{^h _{ijk}} $, along with the field equations,one finds that in a conformally flat $ (\ts C{^h _{ijk}}) = 0 $ background space time  the above equation \eqref{transp1} reduces to 
\begin{align}\label{transp3}
(\nabla_l + \theta /2) (a_+ ^{(1)})+ \frac{R}{3}((a_+ ^{(0)}) = \frac{1}{2} e_+ ^{ij}[4 \nabla_i \nabla^c \delta_j ^d - \delta^c _i \delta^d _j\nabla^2 - h^{cd}\nabla_j \nabla_i)](v_{0 cd}).
\end{align}
which exhibits the possibility of the background curvature R and the nonlinear derivatives of the primary amplitudes $ v_0 $ possibly influencing the transport of $ v_1 $, the correction to the primary amplitude.
Instead of a perfect fluid if one has a dissipative fluid with the energy momentum tensor 
 \begin{equation}
 T_{ij} = (\rho + p)u_i u_j + p g_{ij} - 2\eta \sigma_{ij} -\zeta\theta h_{ij},
 \end{equation}
where, apart from the usual definitions of $ p, \rho, u^i, $ one has $ \eta, \zeta, \sigma_{ij},\theta $ representing the shear viscosity, the bulk viscosity, the shear tensor, and the scalar of expansion, respectively then the perturbed field equations\cite{arp99}
\begin{equation}
\hat R_{ij} = \kappa [\hat T_{ij} - \frac{1}{2}( g_{ij} \hat T + \hat g_{ij} T)]
\end{equation}
give, after using the gauge condition $ \hat g_{ab}u^b = 0, $ along with the fact that the unperturbed streamlines are geodesics, the set of equations \begin{equation}
\ts H{^{ij} _{ab}}\hat R_{ij} = (\kappa/2)[\hat g_{ab} (\rho -p + \zeta \theta) + h_{ab}(4\zeta \hat \theta/(1+3c_s ^2))-4 \eta \hat \sigma_{ab}],  
\end{equation}   
with 
\begin{align}
&\ts H{^{ij}  _{ab}} = h^i _a h^j _b - \alpha h_{ab} u^i u^j,\\ 
&\hat \theta = \hat u^k _{, k}+\frac{g^{ka}}{2} (g_{ka,b} \hat u^b + \hat g_{ka,b}u^b) + \frac{\hat g^{ka}}{2} g_{ka,b} u^b, \\
&\nabla_j \hat u_i = \hat u_{i,j} - \frac{u^b}{2} (\hat g_{ib,j} + \hat g_{jb,i}- \hat g_{ij,b}) + \{ ij,b \} (u_k \hat g^{kb} + \hat u^b).       
\end{align}
Applying the high frequency approximation and associated ansatz and equating the coefficients of $ \epsilon $ terms, the leading order $ \epsilon^{-2} $ yields the dispersion relation as earlier, with its determinant being referred to the tetrad $ (u^a, k^a, e^a _1, e^a _2) $
 \begin{equation}
 l^4 \omega^6 [\omega^2 - c_s ^2 k^2] = 0,
 \end{equation}
 which gives, as in the case of perfect fluid distribution, the three modes $ l^2 = 0,$ corresponding to the gravitational waves,moving along the null rays with $ T^a = l^a, $ the sound wave mode $ H = \frac{1}{2} [c_s^2 h^{ab} - u^a u^b)l_a l_b $, with rays having the tangent $T^a = \omega(c_s k^a/k + u^a) $ and the matter mode $ H = u^a l_a $, with tangent vector $ u^a. $ \\
 If one now considers the quasi-parallel transport of  $ e^a _1  $ and $ e^a _2 $ as defined in \cite{epb87}, and simplifies the transport equation for the primary amplitudes, one gets the simple relation,\cite{arp99}\begin{equation}
[l^i\nabla_i + \frac{1}{2} \nabla_i l^i +\kappa\eta\omega] e_+ ^{ab} f_{ab} = 0 \Rightarrow [\nabla_l + \frac{1}{2} \nabla_i l^i + \kappa\eta\omega)\begin{pmatrix}
a^{(0)} _+ \\
a^{(0)} _\times \end{pmatrix} = 0,
\end{equation}
an equation similar to the one with perfect fluid but with an extra term  proportional to the viscosity $ \eta.$\\
 Writing in terms of the total amplitude $ A^2 = 2(\vert a^+ \vert^2 + \vert a^\times \vert^2 ), $  the equation for the amplitude transport in the dissipative fluid for the gravitational waves, comes out to be [Prasanna 99]\begin{equation}
( D + \nabla^i l_i ) A^2 = - 2 \kappa \eta \omega A^2,
\end{equation}
clearly showing the presence of a damping term due to shear viscosity,
which seems to indicate that in the presence of viscosity, the propagation of gravitational waves could be influenced by the medium, a result that requires further investigation.Thus \textit{the detection of gravitational waves could be opening anew  window to look at the unknown Cosmos more effectively}.\\

Finally it is worthwhile to consider how the LIGO project was commissioned and carried out to finally achieve the findings. ``LIGO research is carried out by
the LIGO scientific collaboration (LSC), a group of more than 1000 scientists
of the collaboration.from universities around the U.S. and in 14 other countries. More than 90 universities and research institutes in the LSC develop detector technology and analyze data; approximately 250 students are strong contributing members of the collaboration.The LSC detector network includes the LIGO interferometers and the GEO600 detector.The GEO team includes scientists at the MPI for Gravitation physics (Albert Einstein Institute AEI), Leibnitz Universitat, Hannover alongwith partners at the university of Glasgow, Cardiff university, the university of Burmingham and few other universities in the UK,and the university of Balearic islands in Spain".
  The Indian efforts in the successful detection of gravitational waves, has given stimulus to the project LIGO-India, also known as INDIGO which is a planned advanced gravitational wave observatory to be located in India as part of the world wide network. The project received in principle approval from the government of India in March 2016. LIGO-India is planned as a collaborative project betwen a consortium of Indian research institutes and the LIGO laboratory in the USA, along with its international partners in Australia, Germany and the UK. Thus the Indian scientific community from research institutes and universities,faculty and students (of Physics, Mathematics and Engineering) have a very ambitious goal to look forward to in observing analyzing and creating new science of the Cosmos stimulating both the academics and the intellect.

\end{document}